\def\year{2020}\relax
\documentclass[letterpaper]{article} 
\usepackage{aaai20}  
\usepackage{times}  
\usepackage{helvet} 
\usepackage{courier}  
\usepackage[hyphens]{url}  
\usepackage{graphicx} 
\usepackage{graphicx}  
\usepackage{amsmath}
\usepackage{multirow}

\title{Posting Bot Detection on Blockchain-based Social Media Platform using Machine Learning Techniques}

\author{Taehyun Kim,\textsuperscript{\rm 1}
Hyomin Shin,\textsuperscript{\rm 1}
Hyung Ju Hwang,\textsuperscript{\rm 1}\thanks{Corresponding authors.}
Seungwon Jeong,\textsuperscript{\rm 2}\footnotemark[1] \\
\textsuperscript{1}{Pohang University of Science and Technology}\\
\textsuperscript{2}{University of Bristol}\\
\{taehyun3401,zhainl,hjhwang\}@postech.ac.kr, eugene.jeong@gmail.com}

\begin{document}

\maketitle

\begin{abstract}
Steemit is a blockchain-based social media platform, where authors can get author rewards in the form of cryptocurrencies called STEEM and SBD (Steem Blockchain Dollars) if their posts are upvoted. Interestingly, curators (or voters) can also get rewards by voting others' posts, which is called a curation reward. A reward is proportional to a curator's STEEM stakes. Throughout this process, Steemit hopes ``good'' content will be automatically discovered by users in a decentralized way, which is known as the Proof-of-Brain (PoB). However, there are many bot accounts programmed to post automatically and get rewards, which discourages real human users from creating good content. We call this type of bot a posting bot. While there are many papers that studied bots on traditional centralized social media platforms such as Facebook and Twitter, we are the first to study posting bots on a blockchain-based social media platform. Compared with the bot detection on the usual social media platforms, the features we created have an advantage that posting bots can be detected without limiting the number or length of posts. We can extract the features of posts by clustering distances between blog data or replies. These features are obtained from the Minimum Average Cluster from Clustering Distance between Frequent words and Articles (MAC-CDFA), which is not used in any of the previous social media research. Based on the enriched features, we enhanced the quality of classification tasks. Comparing the $F_1$-$scores$, the features we created outperformed the features used for bot detection on Facebook and Twitter.\footnote{This paper will appear in the proceedings of ICWSM 2021.}
\end{abstract}

\section{Introduction}
Despite the interest in blockchain technology, the usage of the so-called \emph{decentralized application} (DApp) is still limited. Except for transferring and trading cryptocurrencies, one of the most widely used applications is \emph{Steemit},\footnote{%
	https://steemit.com
} a blockchain-based social media platform. Based on a DApp ranking site,\footnote{%
https://www.stateofthedapps.com
} Steemit had ranked first among all DApps for a long time, and it still ranks sixth, and most DApps with high ranks are based on the \emph{Steem} blockchain \cite{steem2017}, on which Steemit also runs. 

On Steemit, authors get \emph{author rewards} in the form of  cryptocurrencies called STEEM and SBD (Steem Blockchain Dollars) if their posts are upvoted. Interestingly, curators (or voters) also get rewards by voting others' posts, which is called a \emph{curation reward}.
A user is an author if she writes a post, and is a curator if she votes a post (including her own posts).
Rewards are proportional to a curator's staked amount of STEEMs, which is called STEEM POWER.\footnote{%
Converting STEEM to STEEM POWER is instant, but the reverse takes 13 weeks.
}
That is, an \emph{upvote} from a user with more STEEM POWER has a higher value. Each vote consumes a \emph{voting power}, which is regenerated as time goes by. There is also a \emph{downvote} which decreases the reward of a post, which is intended to prevent spams and any malicious content.
Throughout this process, Steemit hopes ``good'' content to be automatically discovered by users in a decentralized way, which is called the Proof-of-Brain (PoB).

However, as on other traditional social media platforms such as Facebook and Twitter, there are many bot accounts that post automatically. We call this type of bot a \emph{posting bot}.
Detection of posting bots may be more critical on Steemit than other platforms, because posting bots on Steemit also get rewards, which discourages real human users from creating good content. 
Due to downvoting, bots that spam frequently cannot survive in terms of rewards.
Therefore, posting bots have evolved in a way that they can write more meaningful posts hence appearing like human accounts.

There are many papers that studied bots on traditional social media platforms.
In particular, some studies detected posting bots on Twitter. Twitter is a microblogging site on which users post messages called Tweets. A Tweet has a 140 (or 280 since November 2017) character limit.
Thus, a text in a tweet is short and relatively easy to analyze compared with other social media platforms such as Facebook and Steemit. On Steemit, there is no restriction on the length of a post, but it is limited by the block size, which is currently 64KB. This is quite enough for most social media posts.\footnote{%
Any media files, e.g., pictures, videos, are uploaded to a traditional cloud service, and only links to the media are included in the post.}
Another complication of detecting bots on Steemit is its high level of anonymity because of the decentralized nature of blockchain.
Due to its financial rewards, relatively long texts, high level of anonymity, it is both important and challenging to detect posting bots on Steemit.

To the best of our knowledge, this study is the first to investigate posting bots on a blockchain-based social media platform. Compared with the bot detection on the traditional social media platforms, the features we created have an advantage that they can be obtained without limiting the number and length of posts. We extract the features by clustering distance between the blog data or replies. These features are obtained from the MAC-CDFA (Minimum Average Cluster from Clustering Distance between Frequent words and Articles), which has not been used in any of the previous social media research. This feature shows similarity between blog data by clustering distances between blog data. Based on the enriched features, we enhanced the classification quality. Comparing the $F_1$-$scores$, the features we created outperformed the features used for the bot detection on Facebook and Twitter.

\section{Related Work}\label{sec:related}

Detecting bots on social media platforms has become an important issue with the growth of social media platforms \cite{allcott2017social,ferrara2016rise}. Many researchers have tried to detect bots with machine learning algorithms \cite{abu2019botcamp,chu2012detecting,clark2016sifting,dickerson2014using,santia2019detecting,varol2017online,wang2010detecting}. They have focused on extracting features that represent patterns of behavior of each account. In particular, features that represent the regularity have played a major role. Some researchers computed the similarity of texts posted by each user, and others measured an entropy of time intervals to express regularity of behavior patterns. In a recent study, \cite{li2019incentivized} pointed out the prevalence of a different type of bot from which users buy votes on Steemit. Most of these bots are easily found from their own advertisements or by examining transfer memos that contain the post URL to be upvoted. In contrast, our focus is a posting bot.

There have been several attempts to extract regularity of texts, especially around Twitter \cite{abu2019botcamp,clark2016sifting,wang2010detecting}. They used a method that considers all pairs of tweets, by defining the similarity between two tweets. Similarity between two tweets is defined in many ways. \cite{wang2010detecting} identified if a tweet is duplicated by another, using the Levenshtein distance. \cite{abu2019botcamp} defined the similarity using the Jaccard index of hashtags contained in each text. \cite{clark2016sifting} considered the longest common sequence of two texts. However, Twitter is different from Steemit in terms of the text length.

In this respect, Facebook is a good example to compare with Steemit. Users of both Steemit and Facebook can write long articles. \cite{santia2019detecting} tried to detect social bots on Facebook with six features including the content-based features. However, they did not extract the features that represent pairwise text similarity. Rather, they computed the innovation rate that represents the vocabulary of each account and also is used on a Twitter dataset \cite{clark2016sifting}. In addition, they proposed six features which are used on a Facebook dataset containing the innovation rate.\\
\indent Features related with the behavioral regularity also give important information. \cite{chu2012detecting} proposed an automated account detection algorithm on Twitter, which measures the entropy of tweeting time intervals. Showing the difference in the distributions of the entropy for each type of account, they emphasized that entropy measures are important for detecting automated accounts. In addition, \cite{chavoshi2017temporal} emphasized that it is important to consider temporal data to detect twitter bots. In this regard, we compute entropy from the sequence of the various activities including transfer of an account. 
\newline 
\indent Other studies have tried to extract various types of features. \cite{dickerson2014using} used sentiment scores to design a social bot classifier by applying the Random Forest algorithm, using several features including the sematic metric. Feature importance extracted from Random Forest algorithm has revealed that semantic metrics play an important role in detecting social bots on Twitter. \cite{varol2017online} extracted six different types of features:  metadata of users and friends, tweet content and sentiment, network patterns, and activity time series.
They highlighted the fact that human and bot accounts have diverse behavior patterns and concluded that 8 - 15 percent of accounts on Twitter are social bots. \newline
\indent There are different approaches to detect bots on social media platform.  \cite{cresci2017social,feng2017groupfound,lee2014early} defined different types of similarities used to detect social bots. \cite{cresci2017social} defined the Digital DNA, which is the sequence of behaviors of a user, and computed the similarity of two sequences. 
\cite{lee2014early} considered the similarity of user names. In addition, \cite{feng2017groupfound} defined the similarity of users' relationships. Both applied each similarity to the hierarchical clustering method. \cite{chavoshi2016debot} is a model to detect twitter bot based on the clustering method. They used the lag-sensitive hashing technique and computed the Pearson correlation of posting time series between each user. Some applied anomaly detection methods \cite{castellini2017fake,minnich2017botwalk}. \cite{castellini2017fake} extracted features and applied them to a denoising autoencoder, a deep learning algorithm, and \cite{minnich2017botwalk} employed an ensemble method of anomaly detection. Some studies are based on the graph structure \cite{cao2012aiding,wang2017sybilscar}. Their approaches are based on the Random Walk \cite{cao2012aiding} or Loop Belief Propagation \cite{wang2017sybilscar}. \cite{boshmaf2015integro,el2018supervised,honer2017minimizing} mixed machine learning algorithms with graph based methods. They defined the similarity between users \cite{el2018supervised} or adjusted each edge weight using a machine learning method \cite{boshmaf2015integro,honer2017minimizing}.
For other studies on Steemit, see \cite{thelwall2018can,casadesus2019steemit,jeong2020centralized}.

\section{Feature Generation}\label{sec:feature_generation_chapter}
In the feature generation section, we describe the features used in the classification. The features are divided into four categories. First, we develop the CDFA group that describes the distance between frequently used words and articles. Second, \cite{santia2019detecting} analyzed the social bots on Facebook. Unlike Twitter, one can write a blog post on Facebook with unlimited characters, similar to Steemit. Therefore, we benchmark the features in \cite{santia2019detecting} and call them a Santia-2019 group. Third, \cite{chu2012detecting} classified the accounts in Twitter into human, bot, cyborg using entropy rate, spam detection, and account properties. However, some of the features are not available or not meaningful to detect the posting bots on Steemit. For example, the kind of twitting device or account verification features are not available on Steemit, and spam detection is not meaningful because out of fourteen spammers, only two appear to be posting bots, and the remaining twelve are humans. Consequently, we benchmark the entropy rate and some of the account properties from \cite{chu2012detecting}, and we denote them as Chu-2012 group. Finally, we added more features related to blockchain in order to observe a relation between blockchain and posting bots. We denote them as blockchain-oriented feature group. In this section, we generate four groups of features. We aim to study the difference in the effect of posting bot detection between the features we created and the features in the previous study (Chu-2012, Santia-2019). Also we want to observe the difference in performances between features with and without the blockchain-oriented ones.

\subsection{CDFA Group}
We introduce the new features called the CDFA group to represent the characteristics of words of a given account. First, to develop the new features, we introduce a clustering method that considers a similarity between articles. Clustering Distance between Frequent words and Articles (CDFA) is a method that transforms word data into real values. We consider the frequent words used by an account and measure the distance between the frequent words and the articles written by the account.

For the given data of the $m$ articles written by an account, to extract the frequent words, we split the articles into words with a space. Let $W_j$ be the set of words in the $j$-th article, $W = \bigcup_{1\le j\le m} W_j$ be the set of all words used in the articles, and $w_i$, $1\le i \le n$, be the words in $W$. Further, for each article, we determine whether a word $w_i$ is used or not. Then we obtain the occurrence vectors $V_j$, $1 \le j \le m$, with the length $n$ in which an element in each vector $V_j$ represents the word occurrence in the $j$-th article. We describe the occurrence vectors as follows:
\begin{equation*}
    V_j[i] = \begin{cases}
    1, & w_i \in W_j \\
    0, & w_i \notin W_j
    \end{cases}.
\end{equation*}
Next, we sum up the occurrence vectors and obtain a total occurrence vector $T$. More precisely, for each element in $T$, the value represents the number of articles where the word appears. Among the values in $T$, words having an occurrence value of 10\% or more of the maximum value in $T$ are defined as frequent words $F$ and we obtain a vector $V_{freq}$ of length $n$ that has value 1 on the frequent words, and 0 otherwise:
\begin{equation*}
    V_{freq}[i] = \begin{cases}
    1, & w_i \in F \\
    0, & w_i \notin F
    \end{cases}.
\end{equation*}

After we determine the vector $V_{freq}$, we compute the Euclidean distance between $V_{freq}$ and $V_j$ and define the distance as $d_j$. For the $m$ distances, we cluster them with the Dirichlet Process Gaussian Mixture Model introduced in \cite{rasmussen2000infinite}. In detail, a maximum number of clusters is set to five. Note that some of the clusters may not contain enough data and the clusters are not appropriate to represent the writing patterns. In this case, we choose the clusters such that sizes of the clusters are at least $\frac{m}{5}$, where the denominator comes from the maximum number of clusters. 

Using CDFA, we obtain the clusters of distances. Among them, we choose a cluster that has the minimum average of the distances. Because posting bots tend to write articles with a fixed form, words in the form would be in the frequent words, and the distances between the frequent words and articles with fixed forms would be small. Therefore, we select the cluster with the minimum average of the distance and denote the cluster as MAC-CDFA. From the MAC-CDFA, we extract the mean, variance of the MAC-CDFA, and the number of clusters that have the size at least $\frac{m}{5}$.

\begin{figure}[h]
  \centering
  \includegraphics[width=1\linewidth]{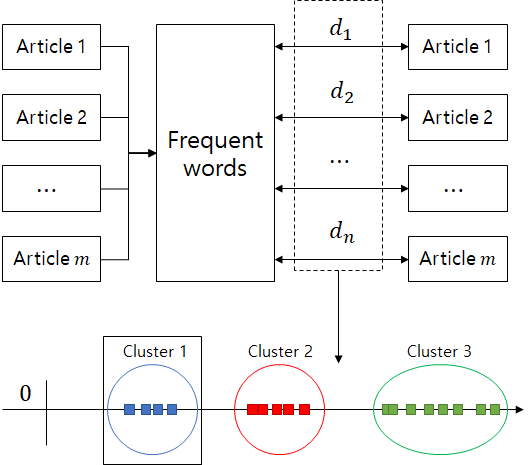}
  \caption{CDFA and MA-CDFA}
  \label{fig:MAC_CDFA_feature_procedure}
\end{figure}

Figure \ref{fig:MAC_CDFA_feature_procedure} shows the procedure of CDFA in brief. From the articles, we extract frequent words and calculate the distance between the frequent words and articles. After that, we cluster the distances and choose the cluster that has the least mean among the clusters.

CDFA can be applied to various datasets. For a blog, there are the title, content, and replies. We applied the CDFA to the title, content, and replies that are written by an account, and we denote them as CDFA-T, CDFA-C, and CDFA-R, respectively. Similarly, we define the MAC-CDFA-T, MAC-CDFA-C, and MAC-CDFA-R. For each MAC-CDFA, we extract three features, thus there are nine features from the CDFA. We call the nine features CDFA feature group (or simply CDFA features). For accounts with less than five blogs, the value of mean and variance of MAC-CDFA-T and MAC-CDFA-C are 0. In addition, the number of clusters via CDFA-T and CDFA-C are 0. Similarly, for accounts with less than five replies, the mean and variance of MAC-CDFA-R and the number of clusters via CDFA-R are 0.

\begin{figure*}[h]
  \centering
  \includegraphics[width=0.8\linewidth]{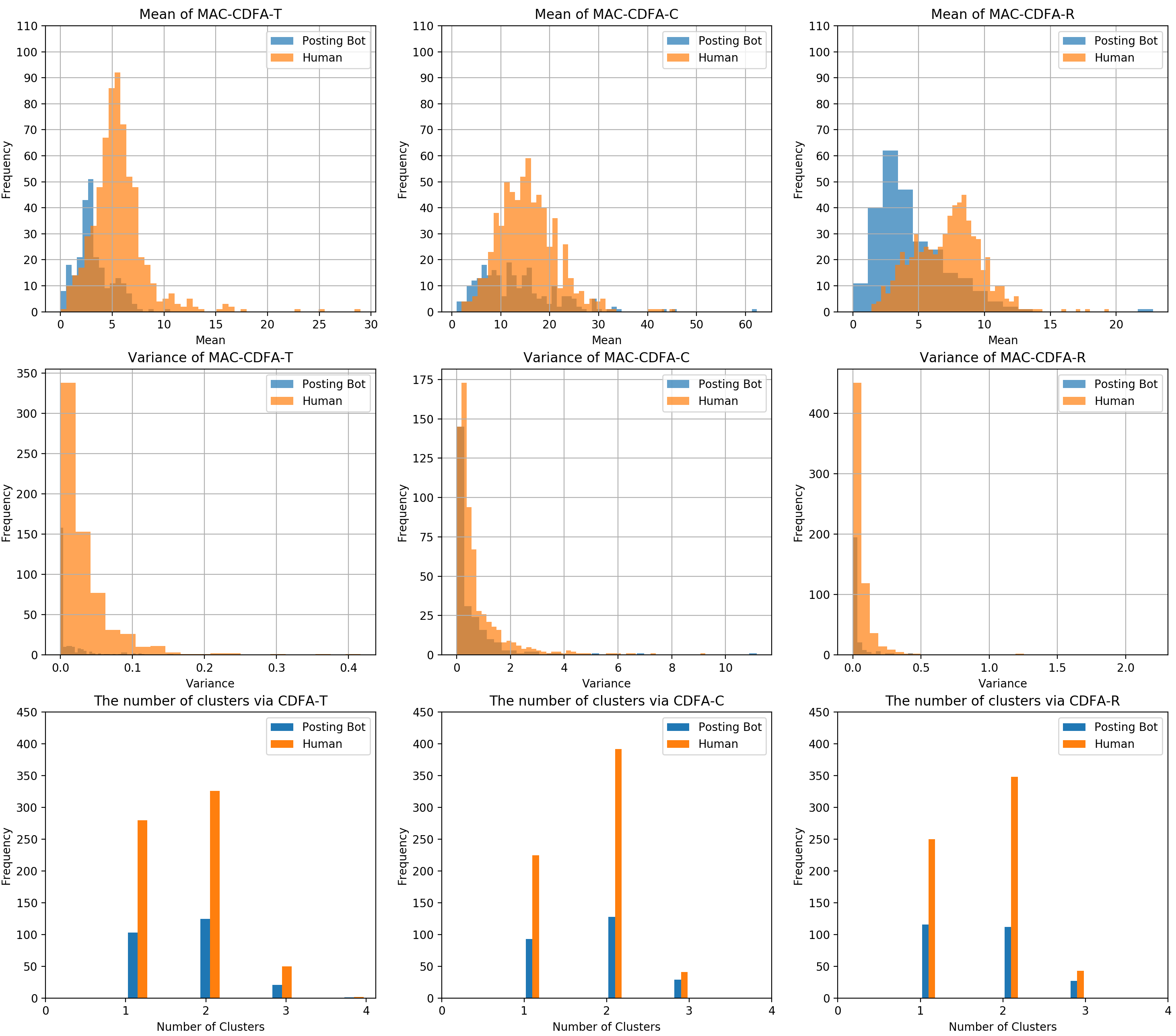}
  \caption{Distribution of Features in CDFA group}
  \label{fig:CDFA_Features_Distribution}
\end{figure*}

Figure \ref{fig:CDFA_Features_Distribution} shows distributions of features in the CDFA group. The top three graphs represent the mean of MAC-CDFA, three graphs in the middle represent the variance of MAC-CDFA, and the bottom three graphs represent the number of clusters via CDFA. To observe meaningful data, we make histograms using accounts with five or more blogs for the left six graphs, and accounts with five or more replies for the right three graphs because the excluded accounts have the value of 0.

\begin{table}[h]
    \centering
  \begin{tabular}{cl}
    \hline
    Index&Feature Name\\
    \hline
    1 & Average of MAC-CDFA-T  \\
    2 & Variance of MAC-CDFA-T \\
    3 & Number of clusters in CDFA-T \\
    4 & Average of MAC-CDFA-C  \\
    5 & Variance of MAC-CDFA-C \\
    6 & Number of clusters in CDFA-C \\
    7 & Average of MAC-CDFA-R  \\
    8 & Variance of MAC-CDFA-R \\
    9 & Number of clusters in CDFA-R \\
  \hline
  \end{tabular}
  \caption{Features in CDFA Group}
  \label{tab:CDFA_features}
\end{table}

In addition to our features, there are numerous other ones related to text similarity and natural language processing. In this section, we will compare the CDFA features with other text-related features: (i) frequent word counts, (ii) term frequency--inverse document frequency (TF--IDF), (iii) Levenshtein edit distance, and (iv) word embedding.

We obtained the frequent word count in the CDFA process. Concurrently, one of the standard methods to analyze text content is TF--IDF, which assigns a weight to each word in a document. In general, in TF--IDF, common words are allotted low weights, whereas uncommon ones have high weights. However, TF--IDF does not represent the features of an account. Assuming that an account frequently uses a word, whereas other accounts scarcely employ it, then the TF--IDF weight of this word will be higher than those of the other words employed in the content written by that account. However, the TF--IDF weight of a word will be relatively low if the word was found in other documents. We calculated the TF--IDF weights of those words occurred for ten times or more.

The Levenshtein edit distance is one of the standard approaches to calculate the distance between two texts. Owing to the long computing time of the model, we sampled the accounts that wrote less than 500 posts and less than 500 replies. Moreover, for a given account, we divided data into titles of blogs, content of blogs, and replies, and calculated the mean of the pairwise distance of each categorized data.

Word embedding is used in natural language processing. To conduct the word embedding, we collected pre-trained dataset in English, German, Spanish, Korean, French, and Russian from Fasttext\cite{joulin2016bag}. In the pre-trained dataset, a word is assigned to a 300-dimensional vector. We denote the set of words in the pre-trained dataset as a \emph{bag of words}. For a given text, we split it into words, and calculate the average of vectors corresponding with the words which are in the bag of words. We denote the average of vectors as a \emph{text vector}. For each account, we calculate a text vector for each blog content or replies, obtain an \emph{account vector} as the average of the text vectors, and use the account vector as a feature. We sampled accounts that the number of words contained in both the bag of words and the words that the account used is 500 or more.

In Table \ref{tab:text_compare}, we compare all the text-related features to the CDFA features. We used Random Forest classifier with Gini index in the classification. The left scores in the table are the $F_1$-$scores$ of the features, and the right scores are the corresponding $F_1$-$scores$ of the CDFA features. Because an account set varies, the corresponding scores of the CDFA features also vary. We observe that the CDFA features outperform the other features.

\begin{table}[ht]
\centering

  \begin{tabular}{|c||c|c|}
    \hline
    Features & Score & Score of CDFA\\
    \hline
    Frequent word counting & 62.78 & 83.72 \\
    \hline
    TF-IDF & 79.38 & 83.72 \\
    \hline
    Levenshtein edit distance & 73.01 & 83.22 \\
    \hline
    Word embedding & 59.05 & 78.14 \\
    \hline
    
\end{tabular}
  \caption{Comparison between text related features and CDFA}
  \label{tab:text_compare}
\end{table}

\subsection{Santia-2019 Group}
In the Santia-2019 group, there are six features; average response time, average comment length, innovation rate, maximum daily comments, number of links and thread deviation.

\subsubsection{Average Response Time}
Steemit users can leave comments on a blog or may leave replies to the comments left on the blog. Moreover, users can leave replies to the replies. We introduce a depth of comments to explain this process. Blogs in Steemit are comments of depth 0. Comments left on blogs are comments of depth 1. If you leave a reply on a comment of depth $n$, then your reply is of depth $n+1$. Then, the comment of depth $n$ you left a reply to is the parent reply to your reply. Response time measures how long each reply has been created since a previous reply was made. Here, the previous reply means a reply written just before the reply among replies whose parent reply is the same. If a reply is the first among them, we compute the time difference from the parent reply. Then, we obtain the response time of each reply. Given a user, average response time is the average of the response times for all replies written by the user.

\subsubsection{Average Comment length}
We generate features related to blogs and replies. One of them is the average comment length. Same as \cite{santia2019detecting}, some of the posting bots generate blog content or replies that are long. We generate the average comment length by averaging lengths of all the blog content and replies written by an account.

\subsubsection{Innovation Rate}
One of the criteria for identifying humans and bots is the diversity of words. To measure the diversity of words, \cite{santia2019detecting} used the innovation rate that represents the decay rate of diversity of words. In \cite{clark2016sifting}, to detect the automation on Twitter, they used the word introduction decay rate $\alpha (n)$. In our case, the whole procedure is the same except the shuffling. Because there are many blogs and replies for some bots, we shuffled the words based on the articles. That means, for $m$ articles, we shuffle the order of the articles, split them with space, and make the sequence of the words. Further, we shuffle three times to obtain the innovation rate.

\subsubsection{Maximum Daily Comments}
Unlike ordinary accounts, bots can write many articles in a day using automated programs. To deal with the bots that generate a massive number of blogs or replies in a short period, we extract the maximum daily comments.

\subsubsection{Number of Links}
We use a regular expression to extract the strings that http or https contain. Even though a regular expression is used, some strings could not be included in the URL in the middle, thus a URL validator is used to filter them out.

\subsubsection{Thread deviation}
This feature represents a regularity in a user's response patterns. We compute response times of all replies left on a blog. Next, we check the average response time corresponding to the blog. Then, for each reply left on the blog, we calculate a difference between the response time of the reply and the average response time of the blog. This difference is called deviation. We calculate the deviation of replies written by a user. Finally, a thread deviation of a user is defined as the average of deviation of replies written by the user.

\begin{table}
\centering
  \begin{tabular}{cl}
    \hline
    Index&Feature Name\\
    \hline
    10 & Average Response Time  \\
    11 & Average Comment length \\
    12 & Innovation Rate \\
    13 & Maximum Daily Comments  \\
    14 & Number of Links \\
    15 & Thread deviation \\
    \hline
  \end{tabular}
\caption{Features in Santia-2019 Group}
\label{tab:Santia-2019_features}
\end{table}

\subsection{Chu-2012 Group}

In Chu-2012 group, there are six features; entropy rate, hashtag ratio, mention ratio, URL ratio, FF ratio and the age of an account.

\subsubsection{Entropy rate}

The entropy rate $\Bar{H}(X)$ is the conditional entropy of an infinite random process $X=\{X_i\}$,
\begin{equation}\nonumber
    \Bar{H}(X) = \lim_{n \to \infty}H(X_n|X_{n-1},\cdots,X_1),
\end{equation}
where the conditional entropy is computed as follows:
\begin{align}
    H(X_n|X_{n-1},\cdots,X_1) = & H(X_1,X_2,\cdots,X_n) - \nonumber \\ & H(X_1,X_2,\cdots,X_{n-1}), \nonumber
\end{align}
and an entropy of a sequence of random variables is defined as
\begin{equation}\nonumber
    H(X_1,\cdots,X_n) = - \sum^n_{i=1}P(X_i = x_i) \log P(X_i = x_i).
\end{equation}
Here, we denote the above equation as an \emph{entropy formula}. Because real data sets are finite, \cite{chu2012detecting} used a corrected conditional entropy, denoted as $CCE$, to estimate the entropy rate. First, they derived the joint probabilities, $P(X_1=x_1,\cdots,X_n =x_n)$, empirically. Then, they computed the conditional entropy based on the empirically derived joint probability. This conditional entropy is denoted by $CE$. Then, they added corrective terms $per(X_n) \cdot EN(X_1)$, where $per(C_n)$ is the percentage of unique sequences of length $n$, and $EN(X_1)$ is the entropy of $X_1$ as follows:
\begin{align}
    CCE(X_n|X_{n-1},\cdots,X_1) = & CE(X_n|X_{n-1},\cdots,X_1)+ \nonumber \\
    & per(X_n) \cdot EN(X_1). \nonumber
\end{align}
They determined $n$ that minimizes $CCE$, and also computed the entropy rate of the sequence of tweeting intervals of each user. We measured the entropy rate of the sequence of comment time intervals and the time difference between comment actions.

\subsubsection{Account Properties}
As we mentioned at the beginning of the feature generation section, some features are available.
In the case of blogs, tag data contains the tags of blogs that represent the main topic of the blogs. Thus, we use a regular expression to extract the hashtags. After obtaining the hashtags, we calculate the \emph{hashtag ratio} by dividing the number of blogs and replies that contain mentions to the number of blogs and replies. We also extract the \emph{mention ratio} in a similar way to the hashtag ratio. In the case of the \emph{URL ratio}, we calculate it via processed data used to extract the number of links by using a similar approach to the hashtag ratio. 

Next, using the follower and following data, we calculate the \emph{FF ratio}.
We obtain the FF ratio by dividing the number of followers by the sum of the number of followers and followings. If the number of followers and followings are 0, the FF ratio is 0. Finally, the \emph{age of an account} is the difference between the time the account was created and the time at the end of the dataset.

\begin{table}
\centering
  \begin{tabular}{cl}
    \hline
    Index & Feature Name\\
    \hline
    16 & Entropy rate  \\
    17 & Hashtag ratio \\
    18 & Mention ratio \\
    19 & URL ratio  \\
    20 & FF ratio \\
    21 & The age of an account \\
    \hline
  \end{tabular}
  \caption{Features in Chu-2012 Group}
  \label{tab:Chu2012_features}
\end{table}

\subsection{Blockchain-Oriented Feature Group}

Based on the blockchain system, we added 12 features, which are listed in Table \ref{tab:Remainder_features}, and term them blockchain-oriented (or simply blockchain) features. In this section, we introduce the blockchain features. \emph{Number of transfers} is the sum of the transfers; \emph{Daily time entropy of transfer} is the entropy setting transfers per day as a random variable in the entropy formula; \emph{Transfer activation time} is the interval between the first and last transfer times; \emph{Daily transfer} is obtained by dividing the number of transfers by the transfer activation time; \emph{In-degree of transfer} of an account is the number of accounts that transferred to the account; \emph{Out-degree of transfer} of an account is the number of accounts that the account transferred; \emph{Entropy of the in-degree accounts} of an account is the entropy setting accounts that are transferred to the account as a random variable in the entropy formula; \emph{Entropy of the out-degree accounts} of an account is the entropy setting accounts that the account transferred as a random variable in the entropy formula, and \emph{Steem-created account} determines whether the account is created by Steem. Initially, to obtain \emph{average transfer per blog or reply}, we calculate the number of blogs or replies and the number of transfers on each day. Subsequently, we divide the number of blogs or replies into the number of transfers on each day and obtain a feature by taking an average. \emph{Average transfer per blog} and \emph{Average transfer per reply} are obtained similarly.

\begin{table}
\centering
  \begin{tabular}{cl}
    \hline
    Index & Feature Name\\
    \hline
    22 & Number of transfers  \\
    23 & Daily time entropy of transfer \\
    24 & Transfer activation time \\
    25 & Daily transfer \\
    26 & In-degree of transfer \\
    27 & Out-degree of transfer \\
    28 & Entropy of the in-degree accounts \\
    29 & Entropy of the out-degree accounts \\
    30 & Steem-created account \\
    31 & Average transfer per blog or reply \\
    32 & Average transfer per blog \\
    33 & Average transfer per reply \\
    \hline
  \end{tabular}
  \caption{Features in the Blockchain-oriented Feature Group}
  \label{tab:Remainder_features}
\end{table}

\begin{figure*}[h]
  \centering
  \includegraphics[width=0.65\linewidth]{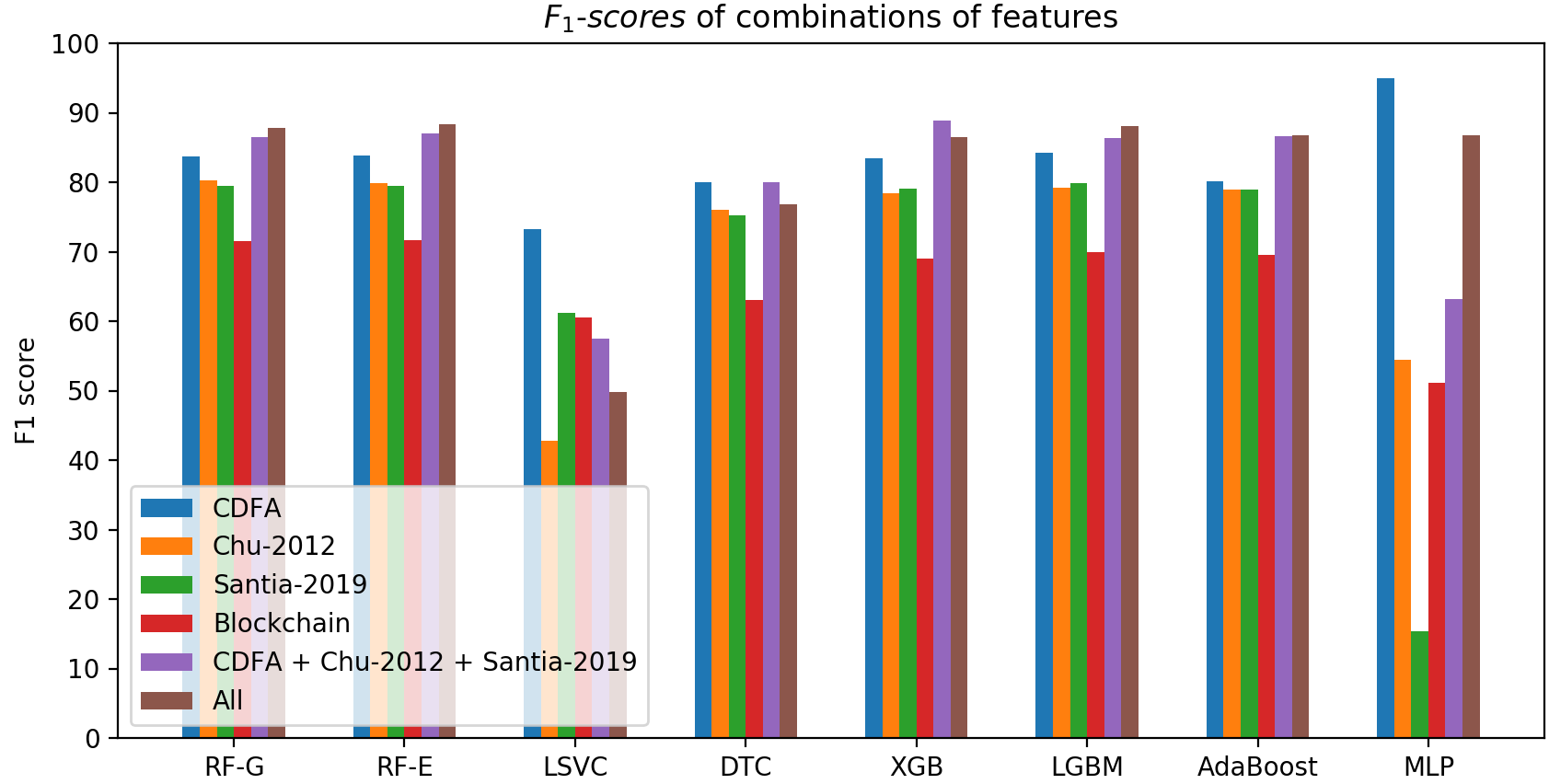}
  \caption{$F_1$-$score$ comparison}
  \label{f1_score_comparison}
\end{figure*}

\section{Posting bot classification}
We explain how to classify the posting bots. First, we introduce a dataset. Second, we clarify an annotation process. Finally, we describe the procedure of classification using several classifiers.

\subsection{Dataset}
Steemit is a social media platform based on the Steem blockchain.
The Steem blockchain is a public blockchain; therefore, all the data are publicly available.\footnote{%
However, there are some types of data that are not stored on this blockchain. First, any media (e.g., pictures, videos, etc.) is stored in a typical centralized cloud service.
Second, broadly, certain activities (e.g., login, logout, and read) that are not publicly available on steemit.com are not stored on the blockchain.
}
Using the data from February 2019 to December 2019, we manually classified humans and bots. A total number of 984 accounts were divided into 325 bot accounts and 659 human accounts. For sampling, we collected the users that write blogs or replies that are 40 times or more. We describe the detailed labeling process in the annotation section.

\subsection{Annotation}
Because Steem is less explored in previous studies, annotation is one of the challenging tasks in our research. Two annotators participated in the annotation, which consists of two stages. In the first stage, both the annotators label the accounts independently using the same dataset. Table \ref{tab:Annotation_first_stage} summarizes the results of the first stage. Subsequently, in the second stage, the annotators compare and discuss their labels. Remark that Cohen's Kappa value is 90.23. Some accounts write several posts or replies like humans, but they are suspicious of using the automated program in some posts or replies. We denote them as \emph{semi-automated accounts}. The annotators analyzed that the Cohen's Kappa value is high because the subjective opinions of two annotators coincide in labeling semi-automated accounts as bots. The annotators agreed to establish the criteria for labeling to deal with the semi-automated accounts and consider the disagreement of 4.17\% for all the accounts. 

Basic criteria for labeling bots were established to be general characteristic of a blog and reply data of the accounts, which are labeled as bots from both the annotators. To set the basic criteria, we define the \emph{form}. When multiple texts have the same form, only the numbers, accounts, and links in the texts change, and the rest is the same. For example, some accounts have the form of ``You got a [$n$]\% upvote from [account].'' In this case, the changes are only in the number $n$ and/or the account part. In addition, some accounts have the form of a table with the ranking of accounts according to some criteria. In this scenario, changes occur only in the accounts in the table. Second, if an account has several forms and writes repeatedly using them, the account is labeled as a bot. For example, when an account runs a gambling app, it is necessary to set the forms such as the winner, amount won, amount they can bet on, and remained funds for gambling. In summary, our basic criterion is that a bot has a certain form or forms in blogs or replies and writes ten or more times in a row using the form or forms. If the blog content or replies of an account match the basic criterion, the account is labeled as a bot. 

\begin{table}[]
\centering
\begin{tabular}{|c|c|c|c|}
\hline
\multicolumn{2}{|c|}{\multirow{2}{*}{Annotators}} & \multicolumn{2}{c|}{Annotator 1} \\ \cline{3-4} 
\multicolumn{2}{|c|}{}                            & Bot            & Human           \\ \hline
\multirow{2}{*}{Annotator 2}        & Bot         & 283            & 26              \\ \cline{2-4} 
                                    & Human       & 15             & 660             \\ \hline
\end{tabular}

\caption{Annotation in the First Stage. The Cohen's Kappa value is 90.23.}
\label{tab:Annotation_first_stage}

\end{table}

However, there are accounts that our basic criterion may not be adequately applied. Therefore, we establish some exceptions. An account that satisfies one of the following cases is labeled as a human: (i) leaving replies to participate in an event or use a service that a bot cannot participate or use easily, (ii) writing a link related to a game-play live streaming and having ten or more replies which do not satisfy the basic criteria, (iii) reporting one's workout records using workout app that has its own abusing detection system, (iv) posting a personal game app status and having ten or more replies which do not satisfy the basic criteria, and (v) posting pictures and having ten or more replies which do not satisfy the basic criteria. In contrast, an account that satisfies one of the following cases is labeled as a bot: (i) copying news and having less than ten replies; (ii) randomly rearranging short sentences. 

The created CDFA features focus on text similarity. In the labeling process, the basic criteria are related to text similarity. However, our labeling does not entirely depend on text similarity. It also considers the opinions or experiences of users. In addition, the labeling process considers copying the news or other types of bots that do not depend on text similarity.

\subsection{Classification Procedure}\label{subsec:classification_procedure}

For the classification, we used several classifiers. In \cite{santia2019detecting} and \cite{chu2012detecting}, Random Forest classifiers with Gini index and Entropy \cite{breiman2001random}, Linear Support Vector classifier \cite{cortes1995support}, and Decision Tree classifier \cite{breiman2017classification} are used to detect bots. In addition to them, we also used more classifiers based on boosting algorithms such as XGBoost \cite{chen2016xgboost}, LightGBM \cite{ke2017lightgbm}, and AdaBoost \cite{freund1999short}. Also, we applied the multi-layer perceptron (MLP) classifier \cite{windeatt2006accuracy} as a representative neural network. We denote Random Forest classifier with Entropy as RF-E, Random Forest classifier with Gini index as RF-G, Linear Support Vector classifier as LSVC, XGBoost classifier as XGB, Decision Tree classifier as DTC, LightGBM as LGBM.

In case of scores, we used the four traditional measurements;  $Accuracy$, $Precision$, $Recall$, and $F_1$-$score$.

To generate the results, we performed five-fold cross-validation. First, we shuffled our dataset and divided it equally into five sets. Next, we choose the first set as the test set, with the remainder becoming the training set. In the training set, we optimized the hyperparameters for each classification algorithm using a grid search via the five-fold cross-validation to obtain a high $F_1$-$score$. Applying the optimized hyperparameters to the classification algorithms, we obtained the models and fit them to the test set and acquired the results. From the divided sets, we can choose five different test sets. Repeating the above procedure, we realized five different results for each model and obtained the final result for each model by taking the average.

\section{Results and Discussion}
We compare the results obtained by employing the four feature groups, and observe that the CDFA group outperforms the other ones. In addition, we derive the results corresponding to the presence and absence of the blockchain features, to ensure that blockchain-oriented features are effective. To interpret the results, we consider the feature importance of each model and rank the features accordingly. For highly ranked features, we further analyze their characteristics.

\subsection{Results}

Note that we categorized the features into the four feature groups: CDFA, Santia-2019, Chu-2012 and blockchain features. For convenience, Santia-2019 is denoted as S, and Chu-2012 as C. Table \ref{tab:score_table_1} shows the results when only one of the four feature groups is applied and the results when blockchain-oriented features are excluded and included. In Table \ref{tab:score_table_1}, we highlight the best scores among the four feature groups for each classifier and the best scores among the classifiers in the cases of including and excluding blockchain-oriented features respectively. We observe that the CDFA group classifies posting bots better than the other feature groups. In addition, including blockchain-oriented features is more effective in detecting posting bots except for linear support vector classifier, decision tree classifier and XGBoost classifier. Finally, as we see in Table \ref{tab:score_table_1}, Random Forest classifier with entropy gives the best score in $Accuracy$ and $F_1$-$score$, Random Forest classifier with Gini index gives the best score in $Precision$, and AdaBoost classifier gives the best score in $Recall$.

\begin{table*}[h]
\centering

  \begin{tabular}{|c||c|c|c|c|c|c|c|}
    \hline
    Models & Scores & CDFA & Santia-2019 & Chu-2012 & Blockchain & CDFA $+$ S $+$ C & All\\
    \hline
    \multirow{4}{5em}{RF-G} 
    & Accuracy  & \textbf{89.63} & 87.40 & 87.10 & 82.42 & 91.46 & 92.28 \\
    & Precision & \textbf{81.73} & 78.66 & 76.23 & 66.93 & 83.48 & \textbf{85.12}\\
    & Recall    & \textbf{85.94} & 82.25 & 83.35 & 77.32 & 89.81 & 90.81\\
    & $F_1$     & \textbf{83.72} & 80.31 & 79.53 & 71.52 & 86.44 & 87.83 \\ 
    \hline
    \multirow{4}{5em}{RF-E} 
    & Accuracy  & \textbf{89.64} & 87.20 & 87.20 & 82.42 & 91.77 & \textbf{92.68} \\
    & Precision & \textbf{82.10} & 77.74 & 75.58 & 67.51 & 84.50 & 84.77 \\
    & Recall    & \textbf{85.75} & 82.43 & 84.06 & 76.85 & 89.81 & 92.28 \\
    & $F_1$     & \textbf{83.83} & 79.83 & 79.51 & 71.70 & 87.03 & \textbf{88.32} \\
    \hline
    \multirow{4}{5em}{LinearSVC} 
    & Accuracy  & \textbf{83.64} & 57.83 & 75.31 & 74.90 & 64.94 & 63.83 \\
    & Precision & \textbf{67.70} & 50.45 & 63.43 & 58.81 & 71.08 & 54.18 \\
    & Recall    & \textbf{80.56} & 49.62 & 65.99 & 64.29 & 56.23 & 56.98 \\
    & $F_1$     & \textbf{73.22} & 42.76 & 61.21 & 60.59 & 57.45 & 49.77 \\
    \hline
    \multirow{4}{5em}{DTC} 
    & Accuracy  & \textbf{87.50} & 85.67 & 84.96 & 76.83 & 87.40 & 85.67 \\
    & Precision & \textbf{76.30} & 69.73 & 70.19 & 60.27 & 76.54 & 72.34 \\
    & Recall    & \textbf{84.33} & 84.01 & 81.29 & 67.76 & 83.78 & 82.59 \\
    & $F_1$     & \textbf{79.99} & 76.00 & 75.29 & 63.04 & 79.97 & 76.82 \\
    \hline
    \multirow{4}{5em}{XGBoost} 
    & Accuracy  & \textbf{89.64} & 86.38 & 87.09 & 81.40 & \textbf{92.99} & 91.46 \\
    & Precision & \textbf{79.39} & 75.03 & 74.05 & 62.97 & \textbf{85.73} & 83.32 \\
    & Recall    & \textbf{88.02} & 82.57 & 84.92 & 76.92 & \textbf{92.36} & 90.02 \\
    & $F_1$     & \textbf{83.47} & 78.36 & 79.06 & 69.06 & \textbf{88.88} & 86.48 \\
    \hline
    \multirow{4}{5em}{LightGBM} 
    & Accuracy  & \textbf{89.94} & 87.09 & 87.50 & 81.91 & 91.36 & 92.48 \\
    & Precision & \textbf{81.60} & 74.98 & 75.34 & 63.89 & 83.23 & 84.49 \\
    & Recall    & \textbf{87.13} & 83.99 & 85.09 & 77.58 & 89.83 & 91.92 \\
    & $F_1$     & \textbf{84.23} & 79.19 & 79.88 & 69.97 & 86.33 & 88.02 \\
    \hline
    \multirow{4}{5em}{AdaBoost} 
    & Accuracy  & \textbf{87.91} & 87.10 & 87.09 & 81.30 & 91.36 & 91.77 \\
    & Precision & \textbf{74.12} & 73.48 & 73.71 & 64.79 & 84.61 & 81.79 \\
    & Recall    & \textbf{87.33} & 85.33 & 85.17 & 75.64 & 88.91 & \textbf{92.50} \\
    & $F_1$     & \textbf{80.09} & 78.89 & 78.94 & 69.58 & 86.61 & 86.70 \\
    \hline
    \multirow{4}{5em}{MLP} 
    & Accuracy  & \textbf{85.77} & 70.93 & 60.98 & 70.94 & 76.93 & 75.81 \\
    & Precision & \textbf{72.60} & 53.54 & 22.76 & 50.54 & 62.77 & 63.94 \\
    & Recall    & \textbf{82.54} & 60.65 & 37.01 & 63.84 & 70.09 & 65.19 \\
    & $F_1$     & \textbf{94.97} & 54.50 & 15.40 & 51.19 & 63.21 & 63.74 \\
    \hline
\end{tabular}
  \caption{Scores of the Models}
  \label{tab:score_table_1}
\end{table*}

\subsection{Feature Importance}
The tree-based ensemble models (Random Forest, Decision Tree, XGBoost, LightGBM, AdaBoost) provide the feature importance. For a tree-based model, the classification is done based on the features in the data set. Feature importance provides information on how the features contributed to improve scores. In the classification procedure, we obtain six different tables of feature importance from the six classifiers. Because we calculate the $F_1$-$score$ by averaging the $F_1$-$scores$ of the five different test sets for each model, we also calculate the feature importance by averaging the importance of the five different train sets. 

\begin{figure*}[htbp]
  \centering
  \includegraphics[width=0.95\linewidth]{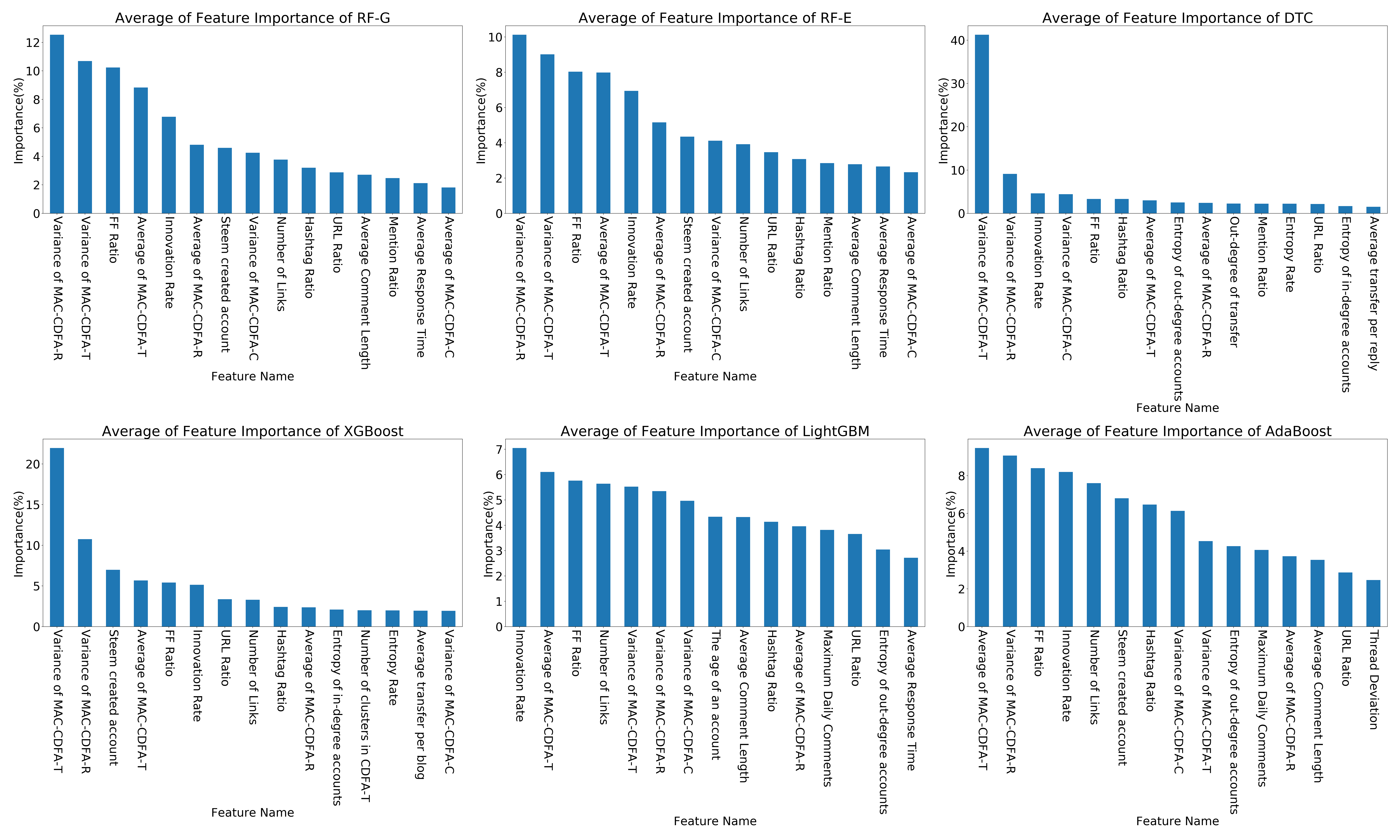}
  \caption{Top 15 feature importance of classifiers}
  \label{feature_importance_graphs}
\end{figure*}

Figure \ref{feature_importance_graphs} shows the results of the top 15 feature importance of each model. The $x$ axis represents the feature names. We observe that some of features in the CDFA group have high importance. 
As we see in Figure \ref{feature_importance_graphs}, only the qualitative analysis on the feature rank is available due to the scattered graphs. To determine the overall ranking, we consider the Borda count in \cite{borda1784memoire} that is one of the popular election methods. The Borda count changes the rank of the relative points and determines the final rank by summing the relative points. There are many methods that change the rank of the relative points. In this study, we use the Dowdall system that calculates the relative points with the reciprocal of the rank. For the average of the feature importance of each model, we determine the rank with respect to the importance. Furthermore, we get the reciprocal of the ranks and sum them up. Finally, we obtain the sum of the relative points of the features and determined the rank. Table \ref{tab:feature_ranking} shows the top five features that have relative points higher than 1. We observe that three of them are in the CDFA group. The FF ratio and the innovation rate are also in Table \ref{tab:feature_ranking}. We will analyze the features in Table \ref{tab:feature_ranking} further in the next section.

\begin{table}[h!]
\centering
  \begin{tabular}{cl}
    \hline
    Rank & Feature Name \\
    \hline
    1  & Variance of MAC-CDFA-R \\
    2  & Variance of MAC-CDFA-T \\
    3  & Mean of MAC-CDFA-T \\
    4  & Innovation Rate \\
    5  & FF Ratio  \\
  \hline
  \end{tabular}
  \caption{Rank of Features (Top 5)}
  \label{tab:feature_ranking}
\end{table}

\subsection{Feature Interpretation}{\label{feature_interpretation}}

\begin{figure*}[htbp]
  \centering
  \includegraphics[width=0.8\linewidth]{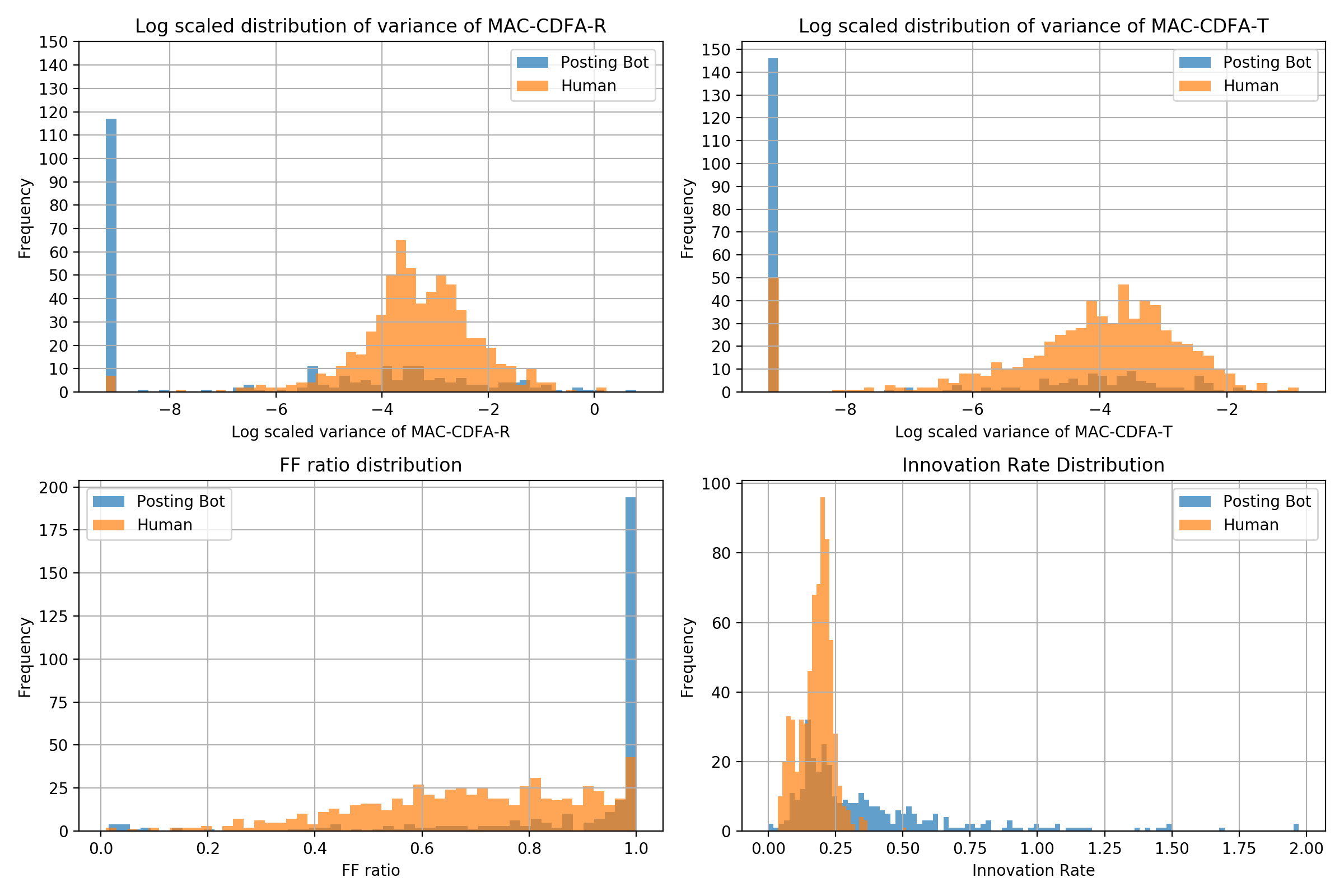}
  \caption{Distributions of important features}
  \label{Distribution_of_important_features}
\end{figure*}

In this section, we check the distributions of important features showed in Table \ref{tab:feature_ranking}, and a feature that has the highest rank among the blockchain features.
Top left graph in the Figure \ref{fig:CDFA_Features_Distribution} shows the distribution of the mean of MAC-CDFA-T for active users who post blogs five times or more. The users in the graph are normally distributed, and the distribution of posting bots has smaller means in the graph. This shows that posting bots tend to post with some forms in titles.\\
\indent The top two histograms in Figure \ref{Distribution_of_important_features} show the log scaled distributions of variance of MAC-CDFA-R and MAC-CDFA-T for active users. We see that the variances of MAC-CDFA-R and MAC-CDFA-T of posting bots are zero more often than humans. In contrast, the log scaled distributions for humans resemble the normal distributions. Consequently, we infer that an active user is a posting bot when the variance of MAC-CDFA-T or MAC-CDFA-R is zero.\\
\indent \cite{chu2012detecting} analyzed that automated bots on Twitter follow numerous users, expecting that humans will follow them in return. However, this scenario is reversed on Steemit. The lower left graph in Figure \ref{Distribution_of_important_features} presents the distribution of the FF ratio. We observe that FF ratios of various posting bots are close to 1 each. This suggests that most of the posting bots do not follow other users. In contrast, humans follow other users actively.\\
\indent A user who has a limited vocabulary has a high innovation rate, whereas a creative user has a low innovation rate. This is well-illustrated in the lower right in Figure \ref{Distribution_of_important_features}. This distribution displays that users with high innovation rates are posting bots, whereas those with low innovation rates are humans.

In contrast, the feature with the highest-ranking among the blockchain features is the out-degree of transfer, and it is ranked 12th. Analysis of this feature demonstrated that 39 accounts had an out-degree of transfer of more than 200, of which 92.3\% or 36 accounts were bots, and 7.7\% or three accounts were humans. In addition, 11.1\% of the bots and only 0.5 \% of the humans had an out-degree of transfer of more than 200. Observation of these accounts suggests that they need to transfer tokens to other accounts, such as running games and events, or manage their tokens within the Steem blockchain. Therefore, from the out-degree of transfer feature, we infer that it assists in detecting these types of bots.

\indent Overall, we observed that the behaviors of posting bots are different from those of humans in numerous aspects. Based on the CDFA features, we obtain the information of the texts close to the representative text structure of each account. In fact, our results demonstrate that using a representative text structure is essential to detect posting bots. Considering the innovation rate, bots produce the same texts with little variations and have restricted vocabularies. In addition, we find an extreme distribution of the FF ratio. From the distribution, we infer that developing a relationship with other accounts is not a primary objective of posting bots. Among the blockchain features, the out-degree of transfer is essential in classification, and we detect some bots that transfer a large amount of cryptocurrencies for running games or managing their tokens.

\section{Conclusion}
The problem of detecting posting bots is one of the essential issues to avail more rewards to human users and motivate them to generate good content. In this paper, we developed features in a CDFA group to detect the posting bots. The CDFA method is used to find frequent words in articles and to measure the distance between the frequent words and the articles. Note that Steemit users can write blogs or replies without limit of length of words like on Facebook. To analyze the posting bot, it is necessary to deal with a large number of blogs or replies with unlimited length because they can generate many articles in a short period of time. Therefore, we calculate the similarity of articles by transforming the articles into real numbers and using a clustering method that can deal with many blogs and replies. With CDFA, we select the MAC-CDFA among the clusters obtained from CDFA and extract features from MAC-CDFA. To compare the performance of features, we benchmark the features introduced in \cite{santia2019detecting} and \cite{chu2012detecting}, and use the $F_1$-$score$ as a comparison measure. The results show that the features in the CDFA group are more effective than other feature groups. To interpret the results, we calculated the feature importance and its rank and performed further analysis of feature distribution.

There is a limitation of our research. In our labeling process, annotations were rarely proceeded for languages that the annotators were not familiar with. In the future research, we expect that new features are developed and detect other kinds of bots in blockchain-based social media platforms. For example, bid voting bots receive cryptocurrency and upvote posts or replies. Although a list of such bots is available, detecting such bots systematically and using them to improve posting bot detection quality would be of interest. Also, we will be able to improve the results by developing customized CDFA features for each language. Finally, we expect that the CDFA features will be used to detect posting bots on social media platforms other than Steemit. 

\section{Acknowledgement}
H.J. Hwang was supported by the National Research Foundation of Korea (NRF) grant funded by the Korea government (MSIT) (2017R1E1A1A03070105) and by Institute for the Information and Communications Technology Promotion (IITP) grant funded by the Korea government (MSIP) (No.2019-0-01906, Artificial Intelligence Graduate School Program (POSTECH)) and by the ITRC (Information Technology Research Center) support program (IITP-2018-0-01441).

\bibliographystyle{aaai}
\bibliography{posting_bot_detection.bib}

\end{document}